\documentclass{aa}  
\usepackage{graphicx}
\usepackage{txfonts}
\usepackage{natbib}
\bibpunct{(}{)}{;}{a}{}{,}
\begin{document}
   \title{Constraining supernova models using the hot gas in clusters of galaxies}

   \subtitle{}

   \author{J. de Plaa \inst{1,2} \and
     	   N. Werner \inst{1} \and
	   J. A. M. Bleeker \inst{1,2} \and
	   J. Vink \inst{2} \and
	   J. S. Kaastra \inst{1} \and
	   M. M\'endez \inst{1} 
	   }

   \offprints{J. de Plaa, \email{j.de.plaa@sron.nl}}

   \institute{
    		SRON Netherlands Institute for Space Research, Sorbonnelaan 2, 3584 CA Utrecht, The Netherlands\\
        	\email{j.de.plaa@sron.nl} \and
		Astronomical Institute, Utrecht University, PO Box 80000, 3508 TA Utrecht, The Netherlands\\
             }

   \date{Received 12 September 2006 / Accepted 18 January 2007}

 
  \abstract
   {The hot X-ray emitting gas in clusters of galaxies is a very large repository of metals produced by supernovae. 
   During the evolution of clusters, billions of supernovae eject their material into this Intra-Cluster Medium (ICM).}
   {We aim to accurately measure the abundances in the ICM of many clusters and compare these data with metal 
   yields produced by supernovae. With accurate abundances determined using this cluster sample we will be able 
   to constrain supernova explosion mechanisms.}
   {Using the data archive of the XMM-Newton X-ray observatory, we compile a sample of 22 clusters. We fit spectra
   extracted from the core regions and determine the abundances of silicon, sulfur, argon, calcium, iron, and nickel.
   The abundances from the spectral fits are subsequently fitted to supernova yields 
   determined from several supernova type Ia and core-collapse supernova models.}
   {We find that the argon and calcium  abundances cannot be fitted with currently favoured supernova type Ia models. 
   We obtain a major improvement of the fit, when we use an empirically modified delayed-detonation model that is 
   calibrated on the Tycho supernova remnant. The two modified parameters are the density where the sound wave in 
   the supernova turns into a shock and the ratio of the specific internal energies of ions and electrons at the shock.
   Our fits also suggest that the core-collapse supernovae that contributed to the 
   enrichment of the ICM had progenitors which were already enriched.}
   {The Ar/Ca ratio in clusters is a good touchstone for determining the quality of type Ia models. 
   The core-collapse contribution, which is about 50\% and not strongly dependent on the IMF or progenitor 
   metallicity, does not have a significant impact on the Ar/Ca ratio. The number ratio between supernova 
   type Ia and core-collapse supernovae suggests that binary systems in the appropriate mass range are very 
   efficient ($\sim$ 5--16\%) in eventually forming supernova type Ia explosions.}


   \maketitle

\section{Introduction}

Clusters of galaxies are the largest gravitationally bound objects in the universe.  About 80\% of the 
baryonic matter in the clusters is in the form of hot X-ray emitting gas that has been continuously enriched 
with metals since the first massive stars exploded as supernovae. The abundances of elements in this hot 
Intra-Cluster Medium (ICM) therefore correspond to the time-integrated yield of the supernova products 
that reached the ICM. About 20--30\% of the supernova products is being locked up in stars in the member galaxies
\citep{loewenstein2004}.
Because of the huge mass of the accumulated metals in the ICM, clusters of galaxies provide a unique way to test
nucleosynthesis models of supernovae on a universal scale.

Since the launch of the ASCA satellite, it has been possible to do abundance studies using multiple elements. 
Several groups \citep[e.g.][]{finoguenov2000,fukazawa2000,finoguenov2001,baumgartner2005} used ASCA observations
of a sample of clusters to study the enrichment of the ICM. They were able to measure the abundances of iron, 
silicon, and sulfur. Also neon, argon, and calcium were sometimes detected, but with relatively low accuracy. The spatial 
distributions of iron and silicon indicated that the core of the clusters is dominated by SN Ia products (Fe), while
the outer parts of the clusters appear to be dominated by core-collapse supernova products (O). Using ASCA 
observations some authors already tried to constrain the specific flavour of supernova type Ia models \citep[e.g.][]{dupke2001}.
Others, like \citet{baumgartner2005} used ASCA data to find that Population-III stars should play 
an important role in the enrichment of the ICM. However, this result is debated \citep{deplaa2006}. 

With the XMM-Newton observatory \citep{jansen2001}, which has both a better spectral resolution and a much larger effective 
area compared to ASCA, it is in principle possible to extend the number of detectable elements to nine. The first abundances
determined from a sample of clusters observed with XMM-Newton were published by \citet{tamura2004}. The general 
picture from the ASCA samples was confirmed, except for the fact that the silicon and sulfur abundance show a 
centrally-peaked spatial profile like iron. The oxygen abundance appears to be more uniformly distributed in 
the clusters.

Recently, \citet{werner2006} and \citet{deplaa2006} analysed deep XMM-Newton observations of the clusters 
\object{2A 0335+096} and \object{S\'ersic 159-03}, respectively. They were able to 
accurately measure the global abundances of about nine elements in the cluster and fit them using nucleosynthesis 
models for supernovae type Ia and core-collapse supernovae. The fits show that $\sim$30\% of all the supernovae in the 
cluster are type Ia and about 70\% are core-collapse supernovae. In their data, they found a clear hint 
that the calcium abundance in these clusters is higher than expected. 

In this paper, we extend the approach of \citet{deplaa2006} and \citet{werner2006} to a sample of 22 
clusters observed with XMM-Newton. We aim to accurately measure the chemical abundances of all robustly 
detected elements and fit model yields of type Ia and core-collapse supernovae to the results. Naturally, 
we discuss the anomalous calcium abundance.         

In our analysis we use H$_{0}$ = 70 km s$^{-1}$ Mpc$^{-1}$, $\Omega_{\mathrm{m}}$ = 0.3, and $\Omega_{\Lambda}$ = 0.7. 
The elemental abundances presented in this paper are given relative to 
the proto-solar abundances from \citet{lodders2003}.

\section{The sample and methodology}

We use XMM-Newton data from a sample of 22 clusters of galaxies in the redshift interval $z$=0--0.2.
The clusters are selected primarily from the HIFLUGCS sample \citep{reiprich2002}, because this sample is well 
studied and contains the brightest clusters in X-rays. We only select the 
observations with the best data quality.

\subsection{Sample selection}
\label{sec:selection}

\begin{table*}[t]
\caption{Summary of the cluster properties of this sample. Data are taken from Table 3 and 4 in \citet{reiprich2002}
apart from the classification.
We list the following properties:
(1) Heliocentric cluster redshift. 
(2) Column density of Galactic neutral hydrogen gas in units of 10$^{20}$ cm$^{-2}$. 
(3) ROSAT flux in the energy range 0.1-2.4 keV in units of 10$^{-11}$ erg s$^{-1}$ cm$^{-2}$. 
(4) Luminosity in the energy range 0.1-2.4 keV in units of $h_{50}^{-2}$ 10$^{44}$ erg s$^{-1}$.
(5) X-ray temperature in keV(note: in this table 90\% errors are used). 
(6) Cluster radius ($R_{500}$) in $h_{50}^{-1}$ Mpc as listed in \citet{reiprich2002}.
(7) Extraction radius used in this analysis in arcmin (0.2$R_{500}$).
(8) Effective XMM-Newton exposure time in ks.
(9) Classification (cooling core [cc] or non-cooling core [non-cc]).
}
\begin{center}
\begin{tabular}{lcccccccccccl}
\hline\hline
Cluster 		& $z$	 & $N_{\mathrm{H}}$ & $f_{\mathrm{X}}$   & $L_{\mathrm{X}}$  &  k$T$		& $R_{500}$   		& $R_{\mathrm{extr}}$	& Exposure& Class\\
			& (1)	 & (2)	 & (3)	   & (4)    &  (5)			& (6) 			& (7)	 	& (8)	  & (9)	\\
\hline
\object{2A 0335+096}	& 0.0349 & 18.64 &   9.16  &  4.79  &  $ 3.01\pm0.07$   	& $1.15\pm0.02$  	& 3.74	 	& 114	  & cc$^{a,b}$	\\ 
\object{A 85}		& 0.0556 &  3.58 &   7.43  &  9.79  &  $ 6.9\pm0.4$   		& $1.68\pm0.06$ 	& 3.42	 	& 12      & cc$^{a,b}$	   \\
\object{A 133}		& 0.0569 &  1.60 &   2.12  &  2.94  &  $ 3.8^{+2.0}_{-0.9}$   	& $1.24^{+0.30}_{-0.16}$& 2.46	 	& 20 	  & cc$^{a,b}$	\\ 
\object{A 1651}		& 0.0860 &  1.71 &   2.54  &  8.00  &  $ 6.1\pm0.4$   		& $1.73\pm0.08$  	& 2.26 	 	& 8       & cc$^{a,b}$	  \\ 
\object{A 1689}		& 0.1840 &  1.80 &   1.45  & 20.61  &  $ 9.2\pm0.3$		& $2.20\pm0.06$  	& 1.31	 	& 36      & cc$^{a,b}$	  \\ 
\object{A 1775}		& 0.0757 &  1.00 &   1.29  &  3.18  &  $3.69^{+0.20}_{-0.11}$   & $1.36\pm0.06$  	& 4.04	 	& 23      & non-cc$^{a,b}$  \\ 
\object{A 1795}		& 0.0616 &  1.20 &   6.27  & 10.12  &  $ 7.8\pm1.0$   		& $1.89\pm0.13$ 	& 3.46	 	& 26      & cc$^{a,b}$	  \\ 
\object{A 2029}		& 0.0767 &  3.07 &   6.94  & 17.31  &  $ 9.1\pm1.0$   		& $2.01\pm0.12$ 	& 2.95	 	& 11      & cc$^{a,b}$	   \\ 
\object{A 2052}		& 0.0348 &  2.90 &   4.71  &  2.45  &  $ 3.03\pm0.04$   	& $1.10\pm0.02$ 	& 3.59 	 	& 29      & cc$^{a,b}$	   \\ 
\object{A 2199}		& 0.0302 &  0.84 &  10.64  &  4.17  &  $ 4.10\pm0.08$   	& $1.43\pm0.04$ 	& 5.38	 	& 23      & cc$^{a,b}$	   \\ 
\object{A 2204}		& 0.1523 &  5.94 &   2.75  & 26.94  &  $ 7.2\pm0.3$   		& $1.82\pm0.05$ 	& 1.32	 	& 19      & cc$^{a,b}$	   \\ 
\object{A 2589}		& 0.0416 &  4.39 &   2.59  &  1.92  &  $ 3.7^{+2.2}_{-1.1}$  	& $1.29^{+0.37}_{-0.22}$& 3.52	 	& 22      & non-cc$^{c}$	   \\ 
\object{A 3112}		& 0.0750 &  2.53 &   3.10  &  7.46  &  $ 5.3^{+0.7}_{-1.0}$   	& $1.53\pm0.14$ 	& 2.29	 	& 22      & cc$^{a,b}$	   \\ 
\object{A 3530}		& 0.0544 &  6.00 &   0.99  &  1.25  &  $ 3.9\pm0.3$   		& $1.47\pm0.14$  	& 3.06	 	& 11      & non-cc	   \\ 
\object{A 3558} 	& 0.0480 &  3.63 &   6.72  &  6.62  &  $ 5.5\pm0.4$		& $1.55\pm0.07$		& 3.70		& 43      & cc$^{a,b}$	   \\
\object{A 3560}		& 0.0495 &  3.92 &   1.52  &  1.60  &  $ 3.2\pm0.5$ 		& $1.14\pm0.12$  	& 2.60	 	& 25      & non-cc$^{d}$	   \\ 
\object{A 3581}		& 0.0214 &  4.26 &   3.34  &  0.66  &  $ 1.83\pm0.04$   	& $0.87\pm0.03$  	& 4.63	 	& 36      & cc$^{a,b}$	   \\ 
\object{A 3827}		& 0.0980 &  2.84 &   1.96  &  7.96  &  $ 7.1\pm1.1$ 		& $2.25^{+0.60}_{-0.37}$& 2.57	 	& 20      & non-cc	   \\ 
\object{A 3888}		& 0.1510 &  1.20 &   1.10  & 10.51  &  $ 8.8\pm1.3$ 		& $2.5\pm0.3$  		& 1.82	 	& 23      & non-cc$^{a,b}$	   \\ 
\object{A 4059}		& 0.0460 &  1.10 &   3.17  &  2.87  &  $ 4.4\pm0.3$   		& $1.40\pm0.06$  	& 3.44	 	& 14      & cc$^{a,b}$	   \\ 
\object{MKW 3S}		& 0.0450 &  3.15 &   3.30  &  2.87  &  $ 3.7\pm0.2$   		& $1.29\pm0.05$  	& 3.24	 	& 35      & cc$^{a,b}$	   \\ 
\object{S159-03}	& 0.0580 &  1.85 &   2.49  &  3.60  &  $ 3.0^{+1.2}_{-0.7}$     & $1.22^{+0.23}_{-0.16}$& 2.45	 	& 113     & cc$^{a,b}$	   \\ 
\hline
\end{tabular}
\begin{minipage}{12cm}
\begin{small}
$^a$ \citet{peres1998}, 
$^b$ \citet{white1997}, 
$^c$ \citet{buote2004}, 
$^d$ \citet{bardelli2002}. \\
\end{small}
\end{minipage}
\end{center}
\label{tab:clusprop}
\end{table*}

Since its launch, XMM-Newton has been used to obtain more than 500 cluster and group observations. However, not all of these 
observations are suitable to use in a sensitive abundance study. We choose the following selection criteria
to ensure that we select a clean and representative sample.

\begin{itemize}
\item The redshift of the cluster is between $z$=0--0.2. We select only local clusters and assume that all
of these clusters have a comparable enrichment history.
\item The cluster core fits inside the field-of-view of XMM-Newton. We need a region on the detector that is 
not heavily 'polluted' with cluster emission to estimate the local background. This excludes large extended nearby 
clusters like Virgo and Coma.
\item The clusters have reported temperatures between $\sim$2 and 10 keV. We exclude groups of galaxies and 
extremely hot clusters.
\item The clusters are part of the HIFLUGCS sample \citep{reiprich2002}.
\item The observation must not suffer from a highly elevated level of soft-protons after flare removal.    
\end{itemize} 

We found 22 clusters that meet these requirements. They are listed in Table~\ref{tab:clusprop}. Together, they
have a total exposure time of 690 ks. Roughly 40\% of the datasets suffer from a high level of so-called residual 
soft-protons. Soft protons have energies comparable to X-ray photons. They can induce events 
in the detector that cannot be separated from X-ray induced events. When the soft-proton flux is high, 
they create a substantial additional background. The soft-proton flare filtering is in general not 
enough to correct for this. An elevated quiescent level is in some cases only detectable as a hard tail 
in the spectrum. We check the spectrum in an 8--11$\arcmin$ annulus centred on the core of the cluster. 
If this spectrum shows an obvious hard tail (more than two times the model count rate at 10 keV), then 
we exclude the cluster from the sample. 

The properties of the clusters we selected are diverse. In Table~\ref{tab:clusprop}, we list a few basic 
properties of the clusters in our sample. The redshifts lie in the range between $z$=0.0214 and $z$=0.184.
Sixteen clusters contain a cooling core (cc).

\subsection{Methodology}

\begin{figure}[t]
\includegraphics[width=\columnwidth]{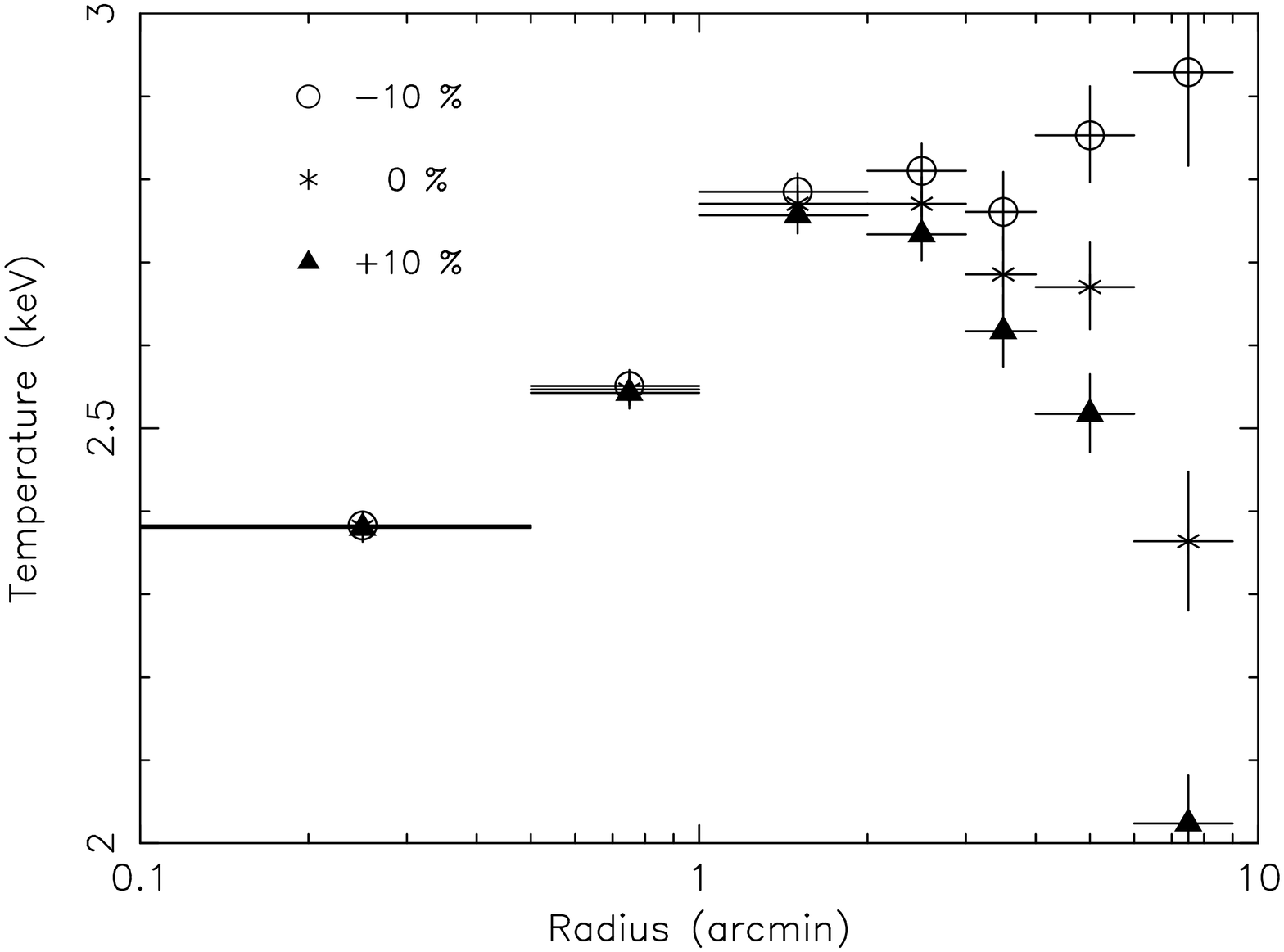}
\caption{Systematic effects in the temperature profile of the cluster of galaxies S\'ersic 159-03. We made 
the three different profiles by subtracting three background spectra with different normalisation. The
blank field backgrounds by \citet{read2003} were scaled up and down by a factor of 10\% and were subtracted 
from the EPIC spectra. The impact to the temperature profile is clear: the temperature is well determined 
in the bright core, but the background subtraction plays an important role in the outskirts.}
\label{fig:systemp}
\end{figure}

Data reduction is done following a procedure that is extensively described in \citet{deplaa2006}. We first 
reprocess the data with SAS version 6.5.0. Then, we filter out soft-proton flares which exceed the 2 sigma 
confidence level. The background subtraction is performed by subtracting a spectrum extracted from a closed 
filter observation that we scale to the instrumental noise level of the particular cluster observation. 
Cosmic background components are included in the spectral fitting phase.

We extract the spectra from a circular region around the core of the cluster. In order to sample comparable
regions in all clusters, we choose a physical radius of 0.2$R_{500}$. The values of $R_{500}$ are taken from 
\citet{reiprich2002}. When we use the radii of 0.2$R_{500}$, we sample the dense core region of the cluster. 
The radii in arcmin for every cluster are listed in Table~\ref{tab:clusprop}.

The spectral components of the background are fitted to a spectrum extracted from an 8--11$\arcmin$ annulus.
This region near the edge of the detector generally contains little cluster emission, 
because we select clusters with a relatively small angular size (see Sect.~\ref{sec:selection}). 
This allows us to estimate the local background without a large bias due to cluster pollution. Small (10\%) 
uncertainties in the background do not affect our analysis. Fig.~\ref{fig:systemp} shows  
that the temperature can be robustly measured in the core of S\'ersic 159-03, even if the background estimate 
would be off by 10\%. Therefore, we concentrate our analysis on the bright cluster cores.   

We fit heuristicly the background spectra with four components: two Collisional Ionisation Equilibrium (CIE) components 
with temperatures 0.07 and 0.25 keV \citep{deplaa2006}, a power-law component with $\Gamma$=1.41, and a 
normalisation fixed to a flux value of 2.24 $\times$ 10$^{-11}$ erg cm$^{-2}$ s$^{-1}$ deg$^{-2}$ 
in the 2--10 keV range \citep{deluca2004}. 
Another CIE component is added to fit any remaining cluster emission. The results of the background fits are then used in 
the fits to the spectra of the core.

For the spectral fitting of the cluster spectra we use the {\it wdem} model \citep{kaastra2004,deplaa2005} which proved 
to be most successful in fitting cluster cores \citep[e.g.][]{kaastra2004,werner2006,deplaa2006}. This model is 
a differential emission measure (DEM) model where the differential emission measure is distributed as a power law 
($\mathrm{d}Y/\mathrm{d}T \propto T^{1/\alpha}$) with a 
high ($T_{\mathrm{max}}$) and low temperature cut-off ($T_{\mathrm{min}}$). We fix $T_{\mathrm{min}}$
to 0.1 times $T_{\mathrm{max}}$ like in \citet{deplaa2006}. We quote the emission-weighted temperature k$T_{\mathrm{mean}}$ 
of the distribution \citep{deplaa2006}.

\section{Results}

\begin{table*}[t]
\caption{Basic properties of the sample of clusters obtained from a fit of MOS and pn data.  
$N_{\mathrm{H}}$ is given in units of 10$^{20}$ cm$^{-2}$, 
$Y = \int n_{\mathrm{e}} n_{\mathrm{H}} \mathrm{d}V$ is in units of 10$^{66}$ cm$^{-3}$, and k$T_{\mathrm{max}}$ and 
$kT_{\mathrm{mean}}$ are in keV. The $\alpha$ parameter is a measure for the slope of the emission-measure distribution.}
\begin{center}
\begin{tabular}{l|cccccc}
\hline\hline
Cluster & $N_{\mathrm{H}}$  & $Y$		& k$T_{\mathrm{max}}$ 	& $\alpha$	& k$T_{\mathrm{mean}}$	& $\chi^2$ / dof \\  
\hline
2A 0335 &  25.71 $\pm$ 0.09  &  18.76 $\pm$ 0.09  &  3.486 $\pm$ 0.016  &  0.360 $\pm$ 0.007  &  2.757 $\pm$ 0.015  &  2927  /  1300  \\
A 85    &  3.25 $\pm$ 0.09   &  30.1 $\pm$ 0.2    &  6.80  $\pm$ 0.18   &  0.50  $\pm$ 0.05   &  5.11  $\pm$ 0.17   &  976  /  739  \\
A 133   &  1.72 $\pm$ 0.10   &  9.25  $\pm$ 0.10  &  4.29 $\pm$ 0.08    &  0.36  $\pm$ 0.03   &  3.40  $\pm$ 0.08   &  1339  /  775  \\
A 1651  &  2.10 $\pm$ 0.16   &  24.0 $\pm$ 0.4    &  7.2 $\pm$ 0.9      &  0.3   $\pm$ 0.2    &  5.9   $\pm$ 1.0    &  857  /  756  \\
A 1689  &  1.96 $\pm$ 0.09   &  88.4 $\pm$ 0.8    &  13.0 $\pm$ 1.1     &  0.52  $\pm$ 0.19   &  9.7   $\pm$ 1.0    &  1106  /  749  \\
A 1775  &  0.48 $\pm$ 0.12   &  9.82 $\pm$ 0.12   &  3.58 $\pm$ 0.19    &  0.03  $\pm$ 0.05   &  3.5   $\pm$ 0.2    &  1124  /  758  \\
A 1795  &  1.08 $\pm$ 0.04   &  40.92 $\pm$ 0.16  &  7.05  $\pm$ 0.13   &  0.55  $\pm$ 0.04   &  5.22  $\pm$ 0.12   &  1703  /  877  \\
A 2029  &  3.23 $\pm$ 0.07   &  72.6 $\pm$ 0.5    &  9.7 $\pm$ 0.4      &  0.51  $\pm$ 0.08   &  7.3   $\pm$ 0.3    &  1279  /  759  \\
A 2052  &  3.22 $\pm$ 0.06   &  5.24  $\pm$ 0.04  &  3.73 $\pm$ 0.04    &  0.411 $\pm$ 0.014  &  2.89  $\pm$ 0.03   &  1336  /  814  \\
A 2199  &  1.16 $\pm$ 0.04   &  11.06 $\pm$ 0.05  &  4.93 $\pm$ 0.08    &  0.32  $\pm$ 0.03   &  3.97  $\pm$ 0.08   &  3328  /  1922  \\
A 2204  &  7.30 $\pm$ 0.13   &  103.8 $\pm$ 0.9   &  10.1 $\pm$ 0.5     &  1.5   $\pm$ 0.4    &  6.5   $\pm$ 0.4    &  982  /  778  \\
A 2589  &  3.54 $\pm$ 0.10   &  5.067 $\pm$ 0.049 &  3.5 $\pm$ 0.3      &  0.00  $\pm$ 0.07   &  3.5   $\pm$ 0.3    &  1087  /  775  \\
A 3112  &  1.12 $\pm$ 0.07   &  24.57 $\pm$ 0.17  &  5.76  $\pm$ 0.11   &  0.44  $\pm$ 0.04   &  4.42  $\pm$ 0.11   &  1221  /  806  \\
A 3530  &  6.1  $\pm$ 0.3    &  2.55 $\pm$ 0.06   &  3.6   $\pm$ 0.4    &  0.03  $\pm$ 0.13   &  3.6   $\pm$ 0.6    &  890  /  709  \\
A 3558  &  3.5  $\pm$ 0.07   &  10.19 $\pm$ 0.06  &  8.1   $\pm$ 0.3    &  0.61  $\pm$ 0.11   &  5.9   $\pm$ 0.3    &  1190  /  768  \\
A 3560  &  3.0  $\pm$ 0.2    &  1.83  $\pm$ 0.04  &  3.3   $\pm$ 0.3    &  0.00  $\pm$ 0.04   &  3.3   $\pm$ 0.3    &  996  /  729  \\
A 3581  &  4.36 $\pm$ 0.11   &  1.85  $\pm$ 0.02  &  2.14 $\pm$ 0.02    &  0.267 $\pm$ 0.008  &  1.765 $\pm$ 0.018  &  1499  /  777  \\
A 3827  &  2.36 $\pm$ 0.12   &  28.2 $\pm$ 0.3    &  8.0   $\pm$ 1.0    &  0.25  $\pm$ 0.19   &  6.7   $\pm$ 1.0    &  1159  /  791  \\
A 3888  &  0.43 $\pm$ 0.12   &  39.0 $\pm$ 0.5    &  9.8   $\pm$ 1.7    &  0.0   $\pm$ 0.2    &  9.8   $\pm$ 2.6    &  1018  /  777  \\
A 4059  &  1.44 $\pm$ 0.06   &  7.85 $\pm$ 0.06   &  4.33  $\pm$ 0.19   &  0.19  $\pm$ 0.06   &  3.7   $\pm$ 0.2    &  1820  /  1426  \\
MKW 3s  &  2.99 $\pm$ 0.06   &  8.35 $\pm$ 0.05   &  4.35  $\pm$ 0.08   &  0.30  $\pm$ 0.03   &  3.53  $\pm$ 0.08   &  1219  /  822  \\
S 159-03&  1.00 $\pm$ 0.03   &  13.7 $\pm$ 0.06   &  3.08  $\pm$ 0.02   &  0.238 $\pm$ 0.010  &  2.59  $\pm$ 0.02   &  3072  /  1495  \\
\hline
\end{tabular}
\end{center}
\label{tab:basic}
\end{table*}

In this section, we apply the {\it wdem} model to the EPIC spectra of the clusters in the sample.
We fix the redshift to the value given in \citet{reiprich2002} and leave the Galactic
neutral hydrogen column density ($N_{\mathrm{H}}$) free in the fit.

\subsection{Basic properties of the sample}

Table~\ref{tab:basic} shows the fit results for $N_\mathrm{H}$ and the temperature structure. The normalisation 
($Y$), the maximum temperature of the DEM distribution (k$T_{\mathrm{max}}$), and the slope parameter $\alpha$ are directly
obtained from the {\it wdem} fit. The emission-weighted temperature (k$T_{\mathrm{mean}}$), derived from the fit parameters,
is a good indicator of the temperature of the cluster core. In our sample these temperatures 
cover a range between roughly 1.7 keV (A3581) and 9.8 keV (A3888). The sample is slightly biased to low-temperature clusters.

A fit with the {\it wdem} model does not always lead to an acceptable $\chi^2$-value (see Table~\ref{tab:basic}). 
This is largely due to systematic errors between the MOS and pn detectors. We describe these systematic differences 
extensively in Sect.~\ref{sec:abun}. A second reason for the high $\chi^2$ can be the complicated temperature structure 
that is often observed in cluster cores. Because 
the {\it wdem} model is just an empirical DEM model, the real temperature distribution in the core of the cluster may 
be somewhat different. Finally, because some weak lines are not yet in the atomic database, small positive residuals 
can arise in line-rich regions like, for example, the Fe-L complex \citep{brickhouse2000}.

\subsection{Abundance determination}
\label{sec:abun}

From each fit to a cluster and from each instrument (MOS and pn), we obtain the elemental abundances of oxygen, 
neon, magnesium, silicon, sulfur, argon, calcium, iron, and nickel. All these abundances, however, can be subject
to various systematic effects.      
We know, for example, that the oxygen and neon abundances are problematic \citep{deplaa2006,werner2006}. The 
spectrum of the Galactic warm-hot X-ray emitting gas (e.g. the local hot bubble) also contains \ion{O}{viii} 
lines that cannot be separately detected with the spectral resolution of EPIC. The brightest neon lines are 
blended with iron lines from the Fe-L complex near 1 keV, which makes an accurate determination of the 
abundance difficult. Therefore, we do not use these two elements in the rest of our discussion.    
In the following sections we discuss two other possible sources of systematic effects and work towards a
robust set of abundances.

\subsubsection{MEKAL vs. APEC}

\begin{table}
\caption{Comparison of the fits using MEKAL and APEC. We fit two CIE models to the spectrum of
S\'ersic 159-03. Normalisation ($Y = \int n_e n_H \mathrm{d}V$) is in units of 10$^{66}$ cm$^{-3}$.}
\begin{center}
\begin{tabular}{lcc}
\hline\hline
Parameter	& MEKAL 	& APEC \\
\hline
$Y_1$		& 5.5$\pm$0.3	& 5.13$\pm$0.11 \\
$Y_2$		& 1.4$\pm$0.3	& 2.20$\pm$0.07 \\
k$T_1$		& 1.99$\pm$0.06 & 1.96$\pm$0.05 \\
k$T_2$		& 4.4$\pm$0.4   & 3.47$\pm$0.08 \\
Si		& 0.315$\pm$0.013&0.331$\pm$0.013 \\
S		& 0.292$\pm$0.019&0.310$\pm$0.019 \\
Ar		& 0.20$\pm$0.05 & 0.20$\pm$0.05 \\
Ca		& 0.45$\pm$0.08 & 0.51$\pm$0.05 \\
Fe		& 0.504$\pm$0.007& 0.514$\pm$0.008 \\
Ni		& 0.63$\pm$0.06 &  0.71$\pm$0.05 \\
$\chi^2 / dof$	& 3882/3091	& 3874/3102 \\
\hline
\end{tabular}
\end{center}
\label{tab:apec}
\end{table}

A possible source of systematic effects is the plasma model that we use.
There is an alternative for the MEKAL-based code called APEC \citep{smith2001}.  
In Table~\ref{tab:apec} we show a comparison of two spectral fits to the spectrum of S\'ersic 
159-03 using a two-temperature model. One fit is performed using the 
MEKAL based CIE model, and the other with APEC. 

In general, the differences between the MEKAL and APEC fits are minor. Both plasma models are 
equally well capable of fitting the spectrum, which is clear from the $\chi^2$ values 
(3882/3091 [MEKAL] and 3874/3102 [APEC]). We note that there is no standard routine
to fit a {\it wdem}-like emission-measure distribution with APEC. Therefore, we test
two-temperature models here, which is often also a good approximation for the temperature
structure in a cluster core. The only differences between the best-fit parameters of the 
two models are in the higher temperature component. MEKAL and APEC find a slightly 
different mix between the high and low temperature component. Despite the small differences in temperature 
structure between the two codes, the derived abundances are consistent within errors.
This conclusion is in line with a similar comparison of the two codes by \citet{sanders2006}.

\subsubsection{Cross-calibration issues}
\label{sec:cal}

\begin{figure}[bt]
\begin{center}
\includegraphics[width=\columnwidth]{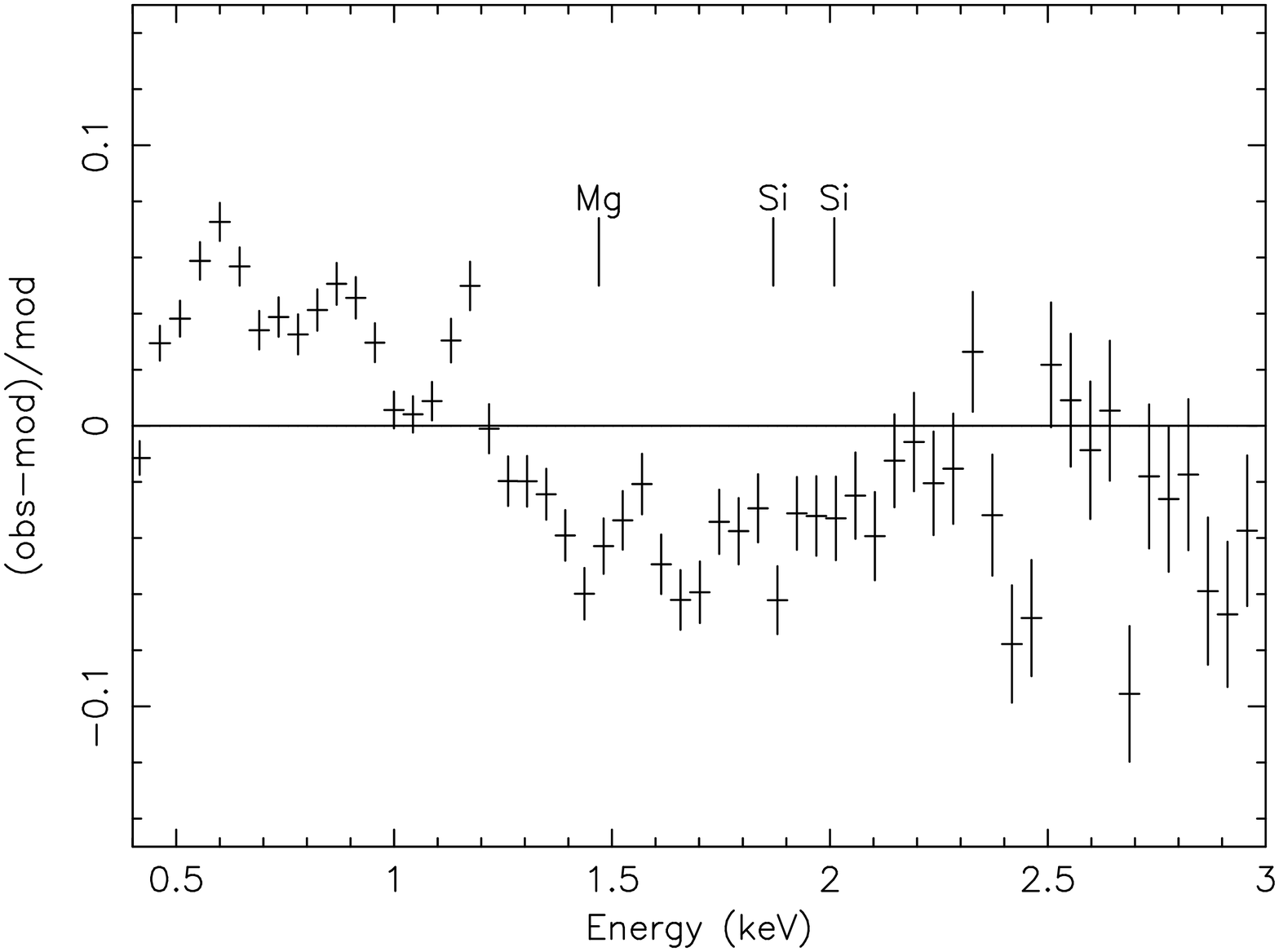}
\end{center}
\caption{Residuals showing the calibration difference between EPIC MOS and pn. We show
the residuals of the cluster S\'ersic 159-03 as a typical example. The data points in
this plot show the difference between the pn data and the best-fit model of MOS: 
(observed[pn]-model[MOS])/model[MOS]. We indicate the (zero-redshift) line energies for the
strongest affected lines in this band: \ion{Mg}{xii} L$\alpha$ (1.47 keV), \ion{Si}{xii} HE 4
(1.87 keV) and \ion{Si}{xiv} L$\alpha$ (2.01 keV).}
\label{fig:calproblem}
\end{figure}

EPIC cross-calibration efforts \citep{kirsch2006} have shown that there are systematic differences 
in effective-area between MOS and pn that are of the order of 5--10\% in certain bands. Systematic 
errors of this magnitude can have a large impact on abundance measurements. We fit the MOS and pn spectra 
separately to investigate the impact on our abundance estimates.

The main differences in calibration between MOS and pn can be found in the 0.3--2 keV band. In Fig.~\ref{fig:calproblem}
we show an example of the differences we observe between the two instruments. The pn instrument shows a 
positive excess with respect to MOS in the 0.3 to 1.2 keV band. Note that the models fit well to the spectra from both 
instruments if the spectra are fitted separately. Between 1.2 to 2.2 keV, the pn gives a lower flux than
MOS. These differences can have a significant effect on the abundances that are measured in this band, like for magnesium
and silicon.  

\begin{figure*}[t]
\begin{center}
\includegraphics[width=0.9\columnwidth]{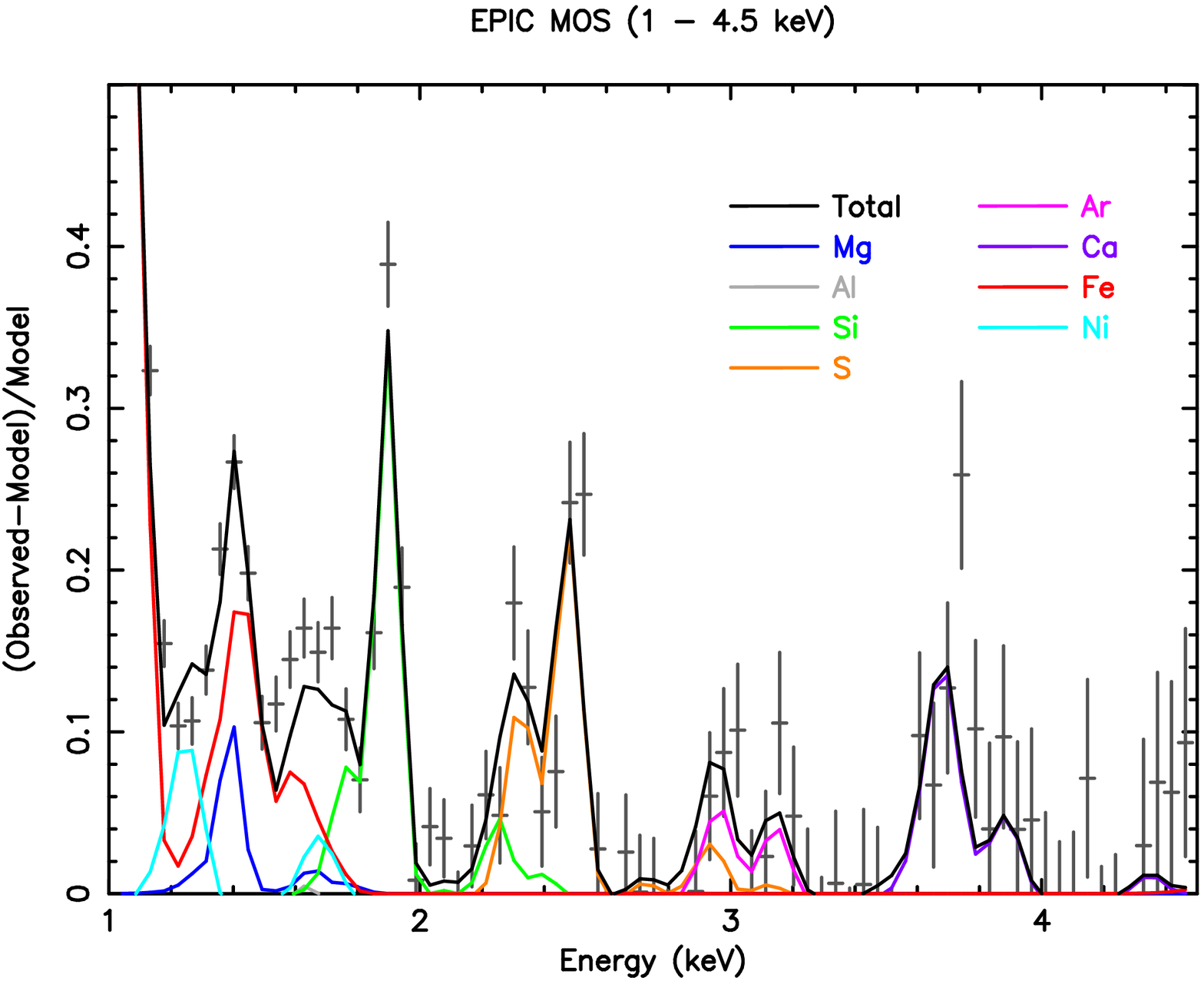}
\hspace{0.2mm}
\includegraphics[width=0.9\columnwidth]{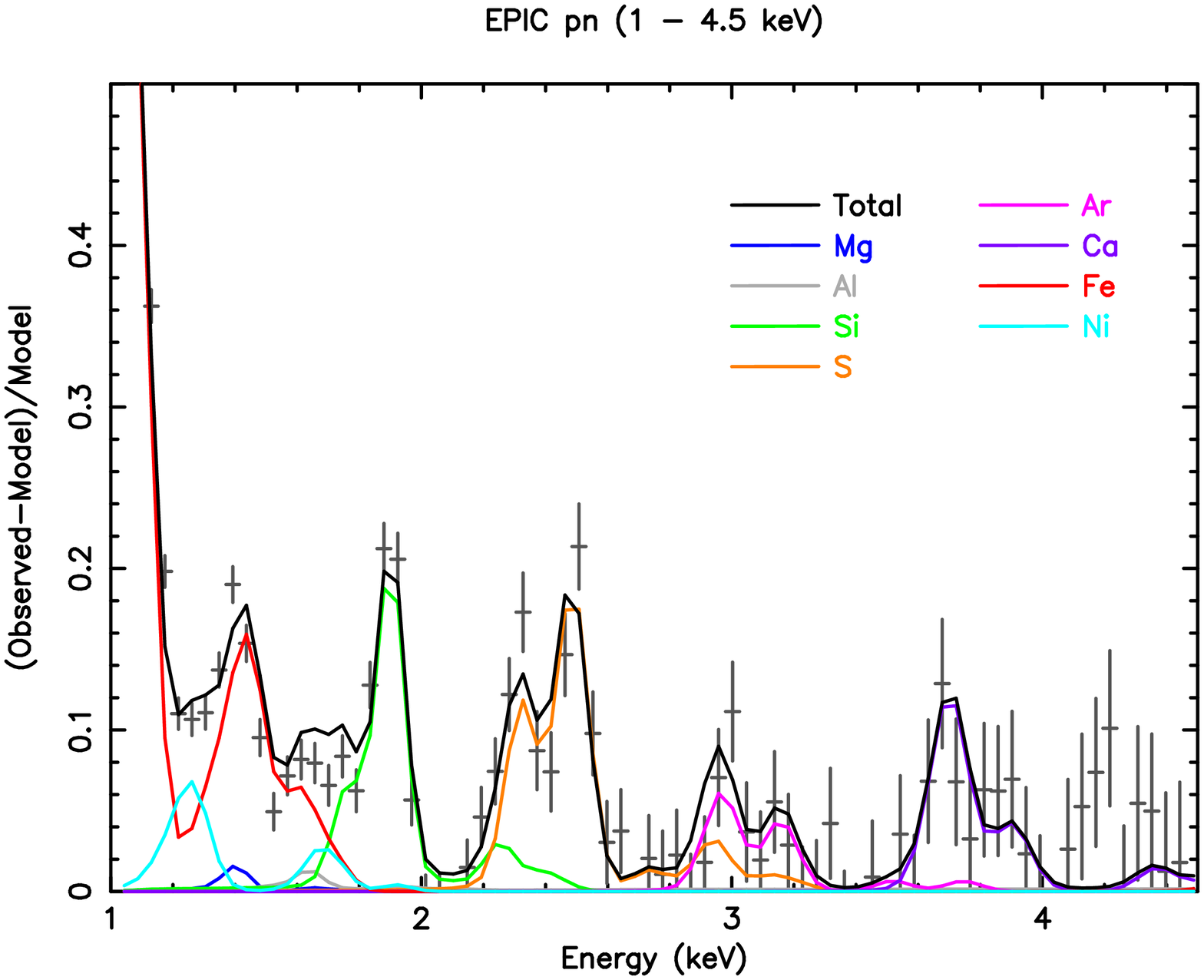}
\end{center}
\caption{Line contributions with respect to the best-fit continuum 
in the EPIC MOS ({\it left}) and pn ({\it right}) spectra of S\'ersic 159-03.}
\label{fig:lines}
\end{figure*}

We show the effect in Fig.~\ref{fig:lines} using the spectrum of S\'ersic 159-03. The plot shows the line 
contributions of the elements that contribute the most to the 1.0--4.5 keV band.
Between roughly 1 and 2.5 keV, the lines in pn appear to contain less flux than their equivalents in MOS, which is in
line with the differences we found between the two instruments. The effect is most notable at 1.4 keV, where 
the magnesium abundance is used by the fit to make up for the difference in flux. Because the
flux of the iron feature at that energy is firmly coupled to the iron-K line complex, the magnesium line flux is
the only one that can fill the gap in flux. The silicon line, however, is quite clean. But there is still a difference 
in flux at the position of this line. At higher energies between 2.5 and 4.5 keV, there is no significant effect anymore, 
which suggests that the sulfur and argon abundances are clean. Above 5 keV, there is a small difference in
the slope that can influence the temperature and subsequently the calcium, iron, and nickel line fluxes.

\begin{figure}[bt]
\includegraphics[width=\columnwidth]{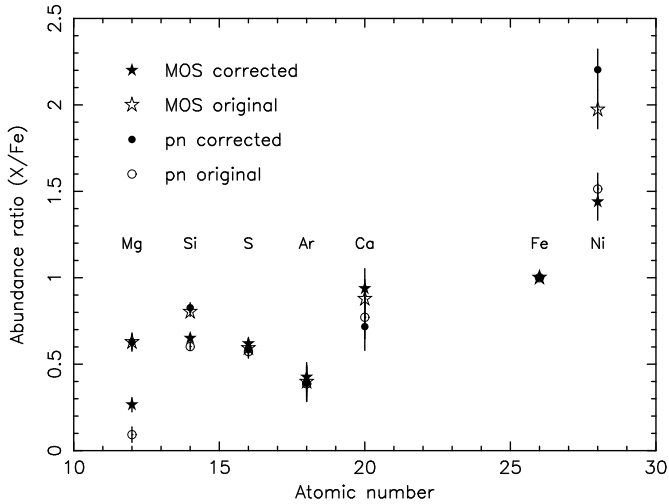}
\caption{Comparison between the abundances measured with the MOS and the pn instruments 
in the cluster S\'ersic 159-03.
We plot the abundance ratios with respect to iron.  The open symbols 
correspond to the original data. The filled symbols show the results when the effective
area is corrected to match the effective area of the other instrument.}
\label{fig:mos-pn}
\end{figure}

Because abundances are directly derived from these line fluxes, we should be able to see the 
differences in the measured abundances. In Fig.~\ref{fig:mos-pn}, we show the abundance ratios 
for MOS and pn separately. We use the data of S\'ersic 159-03 as an example, because it is 
representative for the whole sample. The values for the sulfur, argon, and calcium 
abundances appear to be consistent within errors in the two instruments. However, 
the silicon and magnesium abundances are clearly not. From spectral fits to the pn spectra we obtain 
systematically lower abundances for silicon and magnesium relative to MOS. 

In order to check whether the systematic differences in the abundances are largely due to effective area effects,
we fit the S\'ersic 159-03 spectra again with corrected effective areas. We correct the MOS effective 
area with a simple broad-band spline model such that it nearly matches the pn effective area 
over the whole band. We do the same for pn. The filled symbols in Fig.~\ref{fig:mos-pn} show 
the corrected abundances. The corrected MOS abundance is consistent with the original pn abundance and vice versa.
Therefore, we can conclude that the effective area is the main contributor to the systematic 
differences between abundances determined with MOS and pn.

We choose to use a conservative approach and estimate the systematic error from 
Fig.~\ref{fig:mos-pn}. There are three elements that suffer from systematic effects: magnesium, silicon, 
and nickel. The systematic error in magnesium has such a large magnitude that we can not obtain
a significant value for it. For silicon and nickel we calculate the weighted average and add the  
systematic error linearly to the statistical error. The error should be large enough 
to cover both the MOS and pn results. Using this method, we derive the following systematic 
errors (relative error with respect to the average abundance value): Si ($\pm$11\%) and Ni ($\pm$19\%). 

\subsubsection{Intrinsic scatter}

\begin{figure*}[t]
\begin{center}
\includegraphics[width=0.9\columnwidth]{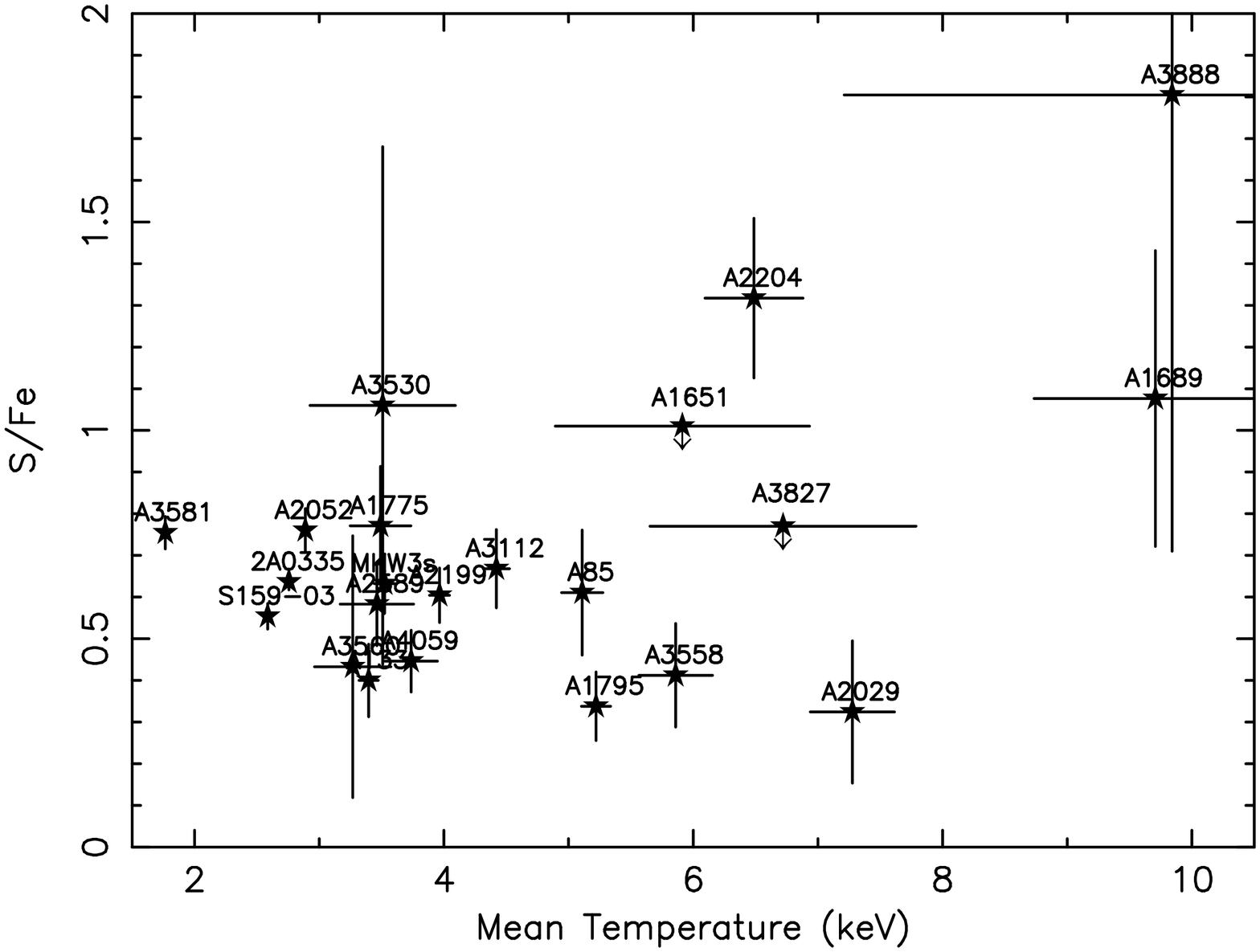}
\includegraphics[width=0.9\columnwidth]{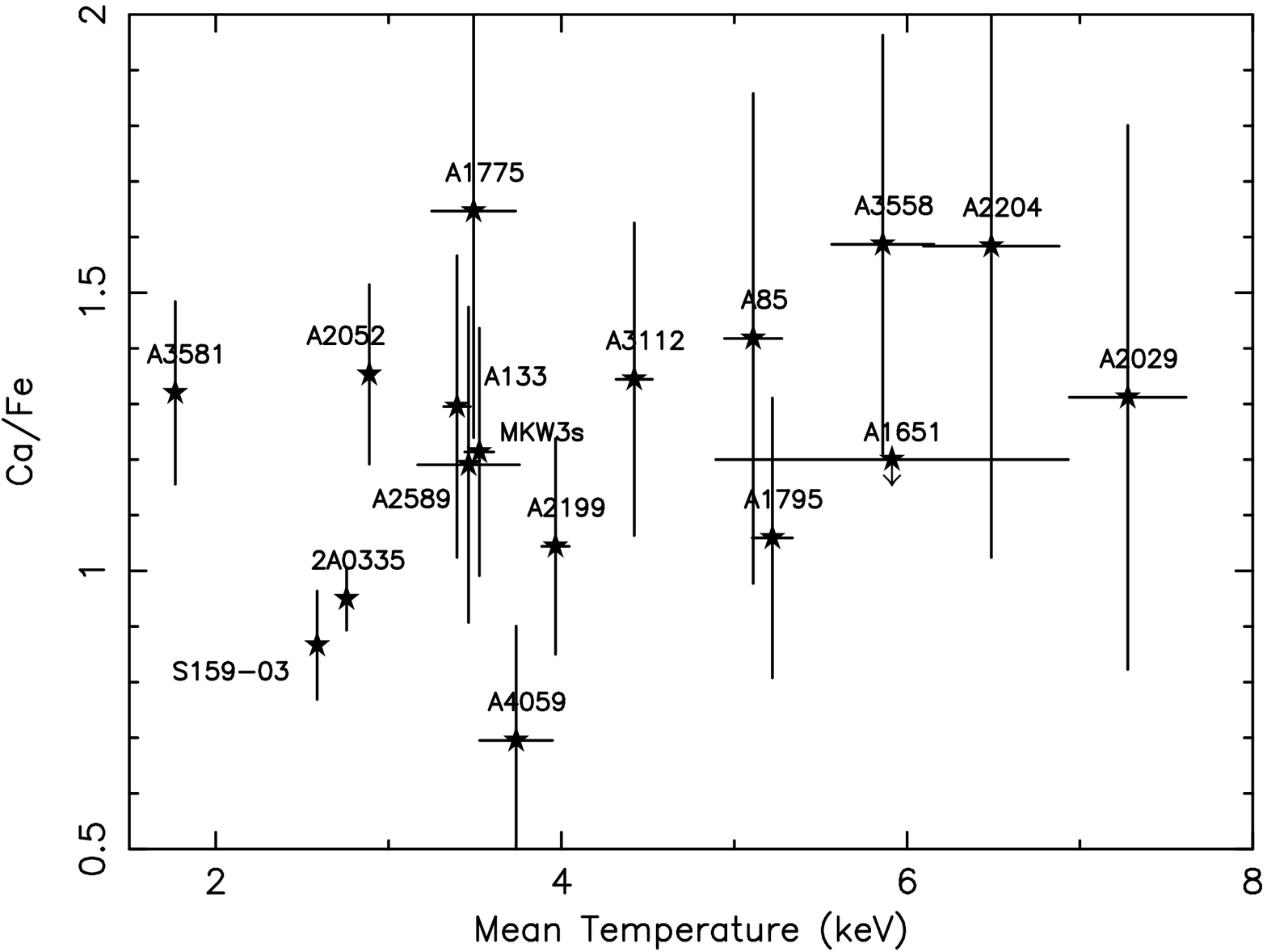}
\end{center}
\caption{Plot of the S-Fe ratio ({\it left}) and the Ca-Fe ratio ({\it right}). In the Ca/Fe plot we do not plot
clusters with 2$\sigma$ upper limits larger than 2.0 for plotting purposes (A1689, A3530, A3560, A3827 and A3888).}
\label{fig:abun}
\end{figure*}

In Fig.~\ref{fig:abun} we show the abundance ratios of S/Fe and Ca/Fe for the individual clusters. 
At first sight, the abundance ratios appear to be consistent with being flat, but they have a small scatter.
In principle we expect to find a scatter, because the clusters in our sample are morphologically different
and may have had different chemical evolution history. The intrinsic differences between the clusters 
need to be taken into account if we can detect the scatter with high significance.

In order to quantify this intrinsic scatter in the abundances, we calculate the error-weighted average of the abundances 
(see Table~\ref{tab:abun}) with an error as described in Eq.~\ref{eq:sigma}:
\begin{equation}
\sigma_{\mathrm{tot}}^2 = \sigma_{\mathrm{m}}^2 + \sigma_{\mathrm{int}}^2
\label{eq:sigma}
\end{equation}
The value for the measured uncertainty ($\sigma_{\mathrm{m}}$) is known from the spectral fits, but the combined uncertainty 
($\sigma_{\mathrm{tot}}$) and the intrinsic scatter in the population of clusters ($\sigma_{\mathrm{int}}$) 
are yet to be determined. We do this by varying
$\sigma_{\mathrm{int}}$ until the $\chi_{\mathrm{red}}^2$ of the weighted average is equal to 1. The variance in the $\chi^2$ distribution for
$n$ free parameters is $2n$ by definition. We use this variance to find the 1$\sigma$ limits on our estimate for 
$\sigma_{\mathrm{int}}$. The values we derive for $\sigma_{\mathrm{int}}$ are listed in Table~\ref{tab:abun}.

We find that the intrinsic scatter ($\sigma_{\mathrm{int}}$) in the silicon and sulfur abundance ratios differs significantly 
from zero. This intrinsic scatter in the data needs to be included in the error of the 
weighted mean. Therefore, we use a new weighted mean for silicon and sulfur with $\frac{1}{\sigma_{\mathrm{tot}}^2}$ 
as weighing factor (see Table~\ref{tab:abun}). Presumably due to lower statistics the $\sigma_{\mathrm{int}}$ of argon, 
calcium, and nickel does not show a significant deviation ($>3\sigma$) from zero, hence we may employ the original 
weighted means.    

\begin{table}
\caption{Weighted averages of the abundance ratios with respect to solar \citep{lodders2003} in our sample. 
Systematic uncertainties due to the effective-area calibration and uncertainties due to intrinsic scatter 
are included in the errors. The $\sigma_{\mathrm{int}}$ that we list here is the intrinsic scatter per data point.}
\begin{center}
\begin{tabular}{lcc}
\hline\hline
X/Fe	& Weighted mean	& $\sigma_{\mathrm{int}}$ \\
	& (incl. $\sigma_{\mathrm{int}}$)& \\
\hline
Si/Fe	& 0.66$\pm$0.13 & 0.17$\pm$0.05 \dag \\
S/Fe  	& 0.60$\pm$0.06 & 0.18$\pm$0.06 \\
Ar/Fe	& 0.40$\pm$0.03 & 0.11$\pm$0.05 \\
Ca/Fe 	& 1.03$\pm$0.04 & 0.12$\pm$0.08 \\
Ni/Fe	& 1.41$\pm$0.31 & 0.2$\pm$0.2 \dag \\
\hline
\end{tabular}
\end{center}
\begin{center}
\dag $\sigma_{\mathrm{int}}$ taken from PN data only.
\end{center}
\label{tab:abun}
\end{table}

\subsubsection{Final abundance ratios}

Now, we have derived values for the most relevant systematic uncertainties that affect our abundances. 
Using the total statistical uncertainty ($\sigma_{\mathrm{tot}}$) and the uncertainty in the effective area, we  
calculate the final abundance values with their errors. This final set of abundance ratios is shown in
Table~\ref{tab:abun}. Silicon and nickel are both dominated by the systematic uncertainty in the effective 
area. The sulfur abundance is dominated by the intrinsic scatter.

\section{Discussion}

We determined the elemental abundances in the core of 22 clusters of galaxies with XMM-Newton. 
Most of the abundances are not consistent with proto-solar abundances \citep{lodders2003}.
The intrinsic scatter in the cluster abundance ratios is between 0--30\%, which is quite small. 
Our sample consists of both relaxed and non-relaxed clusters as well as cooling and non-cooling 
core clusters. The small intrinsic scatter shows that the effects of merging, cooling and temperature
structure on the abundance ratios is limited to 30\% in cluster cores. We do not resolve a clear trend 
of abundances with the presence of a cooling-core. 

It is a well established idea that most of the metals from oxygen up to the iron group are generated by 
supernovae. We construct a few models using elemental yields of supernova type Ia (SNIa) and 
core-collapse supernova (SNcc). This analysis is similar to the one described in \citet{werner2006} 
and \citet{deplaa2006}. 

We try several SNIa yields which we obtain from two physically different sets of supernova models 
\citep{iwamoto1999}, namely slow deflagration and delayed detonation models. The W7 an W70 models 
describe a slow deflagration of the stellar core, while the other models are calculated using 
a delayed-detonation (DD) scenario. WDD2 is the currently favoured SNIa explosion scenario.

For the core-collapse supernovae we use the yields from a recent model by \citet{nomoto2006}.
Note that with SNcc we mean all types of core-collapse supernovae including types II, Ib, and Ic. 
We integrate the yields from the model over the stellar population using an Initial-Mass Function (IMF). 
We perform the calculation following \citet{tsujimoto1995}:
\begin{equation}
M_{i} = \frac{\int_{10 ~ \mathrm{M}_{\sun}}^{50 ~ \mathrm{M}_{\sun}} ~ M_i(m) ~ m^{-(1+x)} ~ \mathrm{d}m}{
        \int_{10 ~ \mathrm{M}_{\sun}}^{50 ~ \mathrm{M}_{\sun}} ~ m^{-(1+x)} ~ \mathrm{d}m},
\label{eq:imf}	
\end{equation}      
where $M_i(m)$ is the $i$th element mass produced in a star of main-sequence mass $m$. We use
a standard model with Salpeter IMF ($x$=1.35) and solar-metallicity (Z=0.02). 

For every element $i$ the total number of atoms $N_i$ is a linear combination of the number
of atoms produced by a single supernova type Ia (Y$_{i,\mathrm{Ia}}$) and type cc (Y$_{i,\mathrm{cc}}$).
\begin{equation}
N_i = a\mathrm{Y}_{i,\mathrm{Ia}} + b\mathrm{Y}_{i,\mathrm{cc}} ,
\end{equation}
where $a$ and $b$ are multiplicative factors of type SNIa and core-collapse supernovae respectively.
The total number of particles for an element can be easily converted into a number abundance.
This reduces to a system of two variables ($a$ and $b$) and six data points
(Si, S, Ar, Ca, Fe and Ni). The fits are independent of the values for the solar abundances,
because they are divided out in the procedure. In essence, we fit the absolute abundances in the cluster. 
In the following sections, we present the ratio of the relative numbers of SNIa with respect to 
the total number of supernovae (SNIa + SNcc) that have enriched the ICM. 

The supernova number ratios that we present, reflect the supernova ratio that is fitted to
the abundances that we measure in the ICM. This does not necessarily correspond to 
the true ratio of supernova explosions in the entire cluster over its lifetime \citep{matteucci2005}. 
Because SNIa explode some time after the initial star burst, there may be a difference between 
the fractions of type Ia and core-collapse products locked up into stars. 
However, this delay time \citep[$\lesssim$3 Gyr, ][]{maoz2004} is probably short with respect to the 
formation time scale of the cluster. Therefore, it is likely that the bulk of the metals formed before 
0.1$t_{\mathrm{Hubble}}$. Because this enrichment timescale is an order of magnitude smaller than the 
Hubble time, the instantaneous recycling approximation \citep{tinsley1980} is presumably a reasonable 
approximation in the case of clusters. Using this approximation implies that we ignore the stellar 
lifetimes and thus the delay with which some chemical elements are released from stars into the ICM 
\citep{matteucci2005}. 

We can make a very rough estimate of the systematic uncertainty that we introduce in our supernova ratio 
when we adopt the instanteneous recycling approximation. Galactic evolution models provide an indication about 
how the O/Fe ratio behaves in and around galaxies \citep[for example, ][]{calura2006}. However, galactic 
models are based on specific assumptions and approximations. Therefore, they also contain systematic uncertainties
that are not well known. The plots in \citet{calura2006}, for example, suggest that the fraction of oxygen that 
is locked up in stars is about a factor of two higher then for iron. If this model is a reasonable representation 
of galactic evolution in clusters of galaxies, then we overestimate our SNIa/(SNIa+SNcc) ratio in clusters with
respect to the true supernova ratio with about 40\% at maximum.

\subsection{Solar abundances}

The abundance ratios of silicon, sulfur, and argon that we derive from the sample are lower 
then proto-solar abundance ratios determined by \citep{lodders2003}.
If we fit a constant to the cluster abundance ratios, we obtain a $\chi^2$ of 418 for 5 degrees 
of freedom. This means that the chemical enrichment in cluster cores differs significantly from 
that in the solar neighbourhood.

In order to compare the supernova ratios in clusters with the solar ratio, we fit the supernova models to a 
constructed dataset with solar abundance ratios that have a nominal error of 5\% on every data point. 
From this fit (with a $\chi^2$/dof of 64/4), we find a supernova type Ia (WDD2) 
contribution in the solar abundance of 0.15 ($\pm$ 0.08), which is actually similar to the value 
for our galaxy found by \citet{tsujimoto1995}. This suggests that the abundances in the Sun are 
probably dominated by core-collapse supernovae that usually produce nearly flat abundance ratios.
However, by adopting the instantaneous recycling approximation we might underestimate the type Ia
fraction for the solar neighbourhood with the same factor that we derived from the galactic evolution 
models by \citet{calura2006}.

\subsection{Supernova type Ia models by Iwamoto et al.}

\begin{table}
\caption{Number ratios of supernovae type Ia over the total number of supernovae derived using  
SN Ia models by \citet{iwamoto1999} and by \citep{badenes2006}. The results are from a fit 
which contained the elements from silicon to nickel. We also list the results of the 
comparison to solar abundances.}
\begin{center}
\begin{tabular}{lrr}
\hline\hline
Model & SNIa/SNIa+SNcc & $\chi^2$/dof \\
\hline
Constant& 		 & 418/5 \\
Solar 	& 0.15 $\pm$ 0.08& 64/4  \\
\hline
W7   	& 0.22 $\pm$ 0.06& 152/4 \\
W70  	& 0.26 $\pm$ 0.07& 104/4 \\
\hline
WDD2	& 0.37 $\pm$ 0.09& 84/4  \\
WDD3	& 0.22 $\pm$ 0.06& 105/4 \\
CDD2	& 0.32 $\pm$ 0.08& 86/4	 \\
\hline
Tycho	& 0.72 $\pm$ 0.17& 26/4 \\
\hline
\end{tabular}
\end{center}
\label{tab:supernova}
\end{table}

\begin{figure}[t]
\begin{center}
\includegraphics[width=\columnwidth]{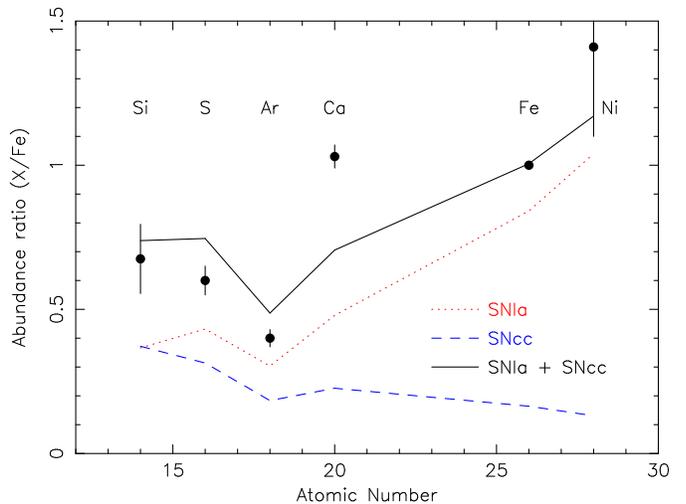}
\end{center}
\caption{Abundance ratios versus atomic numbers for the sample. We fit the supernova yield models
for SN Ia (WDD2) and SNcc ($Z$=0.02 and Salpeter IMF). The black line shows the total fit, while 
the dotted and dashed lines represent the SN Ia and SNcc models respectively.}
\label{fig:wdd2}
\end{figure}

We now try to fit the current supernova models to the data of the sample. In Table~\ref{tab:supernova} we show the 
fit results using the supernova type Ia models W7, W70, WDD2, WDD3, and CDD2. None of the models 
provides an acceptable fit. The model with the lowest $\chi^2$ is the delayed-detonation model WDD2 
($\chi^2$/dof = 84/4). 

The reason why the models fail can be found in Fig.~\ref{fig:wdd2}. The calcium abundance is highly underestimated 
by the models. Moreover, the high calcium abundance forces the fit to increase the SNcc contribution. 
The current models are clearly not able to produce the observed Ar/Ca and Ca/Fe
abundances. This result re-affirms the earlier measurements in 2A 0335+096 \citep{werner2006} and 
S\'ersic 159-03 \citep{deplaa2006}.

\subsection{Supernova type Ia models based on Tycho}

\begin{figure}[t]
\includegraphics[width=\columnwidth]{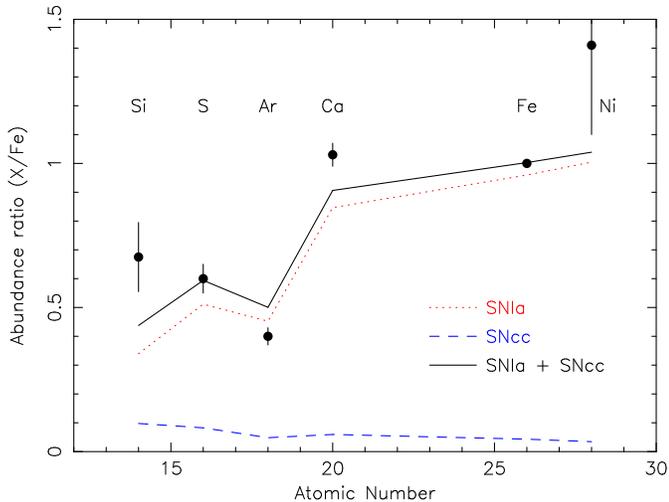}
\caption{Same as Fig~\ref{fig:wdd2}, but now we fit the SNIa yields found in the Tycho supernova
remnant by \citet{badenes2006}. The nickel yield of the Tycho SNIa model was kindly provided by
Carles Badenes (priv. comm.).}
\label{fig:badenes}
\end{figure}

Recently, \citet{badenes2006} compared type Ia models by \citet{chieffi1989} and \citet{bravo1996} 
to XMM-Newton EPIC observations of the \object{Tycho} supernova remnant. The Tycho supernova is thought to have been a 
type Ia supernova. \citet{badenes2006} empirically modified the parameters of their delayed-detonation model to 
fit the Tycho observations. The parameter that mainly determines the outcome of their model is the 
density where the subsonic wave, which runs through the white dwarf during the explosion, turns into 
a supersonic shock. This transition from deflagration to detonation is put in by hand in every DD model.
By modifying this parameter and the ratio between the specific internal energies of ions and electrons ($\beta$), 
they found a best-fit delayed-detonation model that fitted the Tycho observations. 

We take the yields from this best-fit model of Tycho and use them as a supernova type Ia 
model in our fit to the cluster abundances. Note that in Tycho not all the ejected material is 
visible in X-rays, because the reverse shock has not ionised all the material yet. Therefore,
the Tycho results might not reflect the total SNIa yields yet. Despite this caveat, the Tycho 
model provides a major improvement in $\chi^2$ compared to the \citet{iwamoto1999} models 
(see Table~\ref{tab:supernova}). In Fig.~\ref{fig:badenes}, we show that the Tycho model is more successful
in fitting the calcium abundance in clusters. Moreover, the supernova ratios change dramatically, 
the SNIa/(SNIa+SNcc) ratio for this model is 0.72 $\pm$ 0.17. 

This shows that the Ar/Ca and Ca/Fe abundance ratios mainly determine how well   
type Ia models fit. By varying the parameters of the delayed-detonation models,
it is in principle possible to obtain a calcium abundance that fits the observations,
which apparently very effectively constrain type Ia models.

\subsection{Core-collapse models}

In order to test whether our models are also reproducing the core-collapse contribution 
adequately, we need abundances of some typical core-collapse products. Therefore, we estimate the 
oxygen and neon abundance of the sample using the Reflection Grating Spectrometer (RGS) aboard XMM-Newton.
For oxygen, we take the average of the clusters S\'ersic 159-03 and 2A 0335+096,
because these clusters have the highest exposure in our sample and very good RGS data 
\citep{deplaa2006,werner2006}. The O/Fe measurements of the two clusters are not 
statistically consistent possibly due to systematic differences in
the line widths \citep[see][ for an explanation of this effect]{deplaa2006}. Therefore, we
take the average value and assign an error which covers both results within 1$\sigma$. 
The neon abundance is consistent in both S\'ersic 159-03 and 2A 0335+096, hence we take 
the weighted average of the two neon abundances and use them in the rest of the fits.

\begin{figure}[t]
\includegraphics[width=\columnwidth]{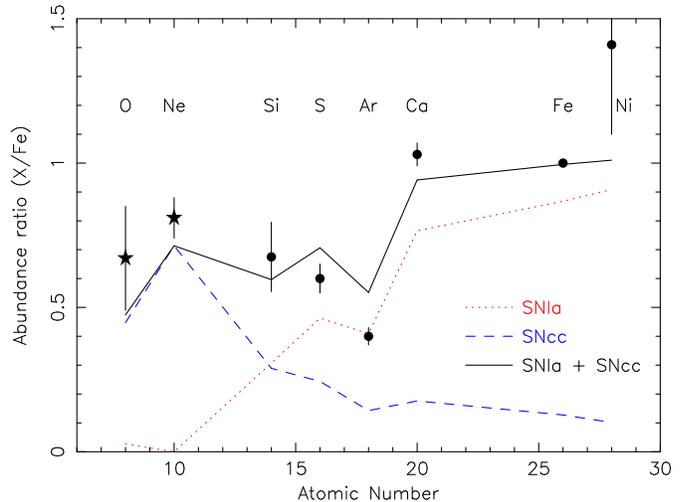}
\caption{Fit using the SNIa yields by \citet{badenes2006}, but now with additional oxygen and neon 
data points (stars) obtained from the RGS spectra of S\'ersic 159-03 and 2A 0335+096. Here, the 
core-collapse model with $Z$=0.02 and Salpeter IMF is used.}
\label{fig:badenes2}
\end{figure}

The fit including O/Fe and Ne/Fe from RGS is shown in Fig.~\ref{fig:badenes2}. We use here a 
standard core-collapse model ($Z$=0.02 and Salpeter IMF) and the type Ia model based on Tycho. The 
trend in O/Ne that is predicted by the core-collapse model, is consistent with the 
O/Ne ratio that we observe. However, the core-collapse contribution needs to increase
with respect to the model used in Fig.~\ref{fig:badenes} to explain the absolute values for 
O/Fe and Ne/Fe (see Table~\ref{tab:core-collapse}). The typical values that we derive are
of the order of 45--60\%. Considering the uncertainties, this number is compatible with
the current supernova type Ia ratio within $z$=0.03 ($\sim$42\%) determined by the 
Lick Observatory Supernova Search (LOSS) \citep{vandenbergh2005}.

In principle, the increase of the core-collapse contribution with respect to the results 
before we included oxygen an neon results in a smaller predicted Ar/Ca ratio. However, 
the data suggest that the Ar/Ca ratio is larger. The plot shows that this particular
core-collapse model still allows a relatively high Ar/Ca ratio, because the absolute 
contributions of silicon, sulfur, argon, and calcium are relatively small in this model.

\subsubsection{Effect of progenitor metallicity on core-collapse yields}

\begin{figure}[t]
\includegraphics[width=\columnwidth]{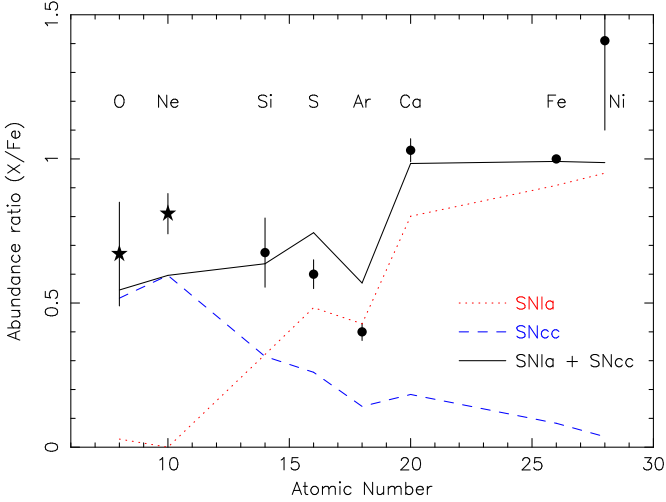}
\caption{Fit using the SNIa yields by \citet{badenes2006} and SNcc yields for Z=0.001 
metallicity progenitors \citep{nomoto2006}.  }
\label{fig:bad_metal}
\end{figure}

\begin{table}[t]
\caption{Results for a supernova fit to the abundance ratios using SNcc models by \citet{nomoto2006} that
are integrated over a Salpeter ($x$=1.35) or ``top-heavy'' IMF ($x$=0). $Z$=0.02 is the solar abundance.
For the type Ia model we used \citet{badenes2006}.}
\begin{center}
\begin{tabular}{lcccc}
\hline\hline
$Z$		& \multicolumn{2}{c}{Salpeter IMF} & \multicolumn{2}{c}{``Top-heavy'' IMF} \\
\hline
		& $\frac{SNIa}{SNIa+SNcc}$ &$\chi^2$/dof &$\frac{SNIa}{SNIa+SNcc}$ &$\chi^2$/dof \\
		&			   &		 & 			&	\\
\hline
0		& 0.55$\pm$0.05	& 79/6		& 0.68$\pm$0.04	& 102/6		\\
0.001		& 0.51$\pm$0.05	& 54/6		& 0.62$\pm$0.04	& 69/6		\\
0.004		& 0.49$\pm$0.05	& 55/6		& 0.62$\pm$0.04	& 58/6		\\
0.02		& 0.44$\pm$0.05 & 40/6		& 0.57$\pm$0.04	& 34/6		\\
\hline
\end{tabular}
\end{center}
\label{tab:core-collapse}
\end{table}

Up to now, we have used a core-collapse supernova model that assumes that the progenitor had a solar metallicity.
\citet{nomoto2006} also provide models where the metallicity ($Z$) of the supernova progenitor is 0, 0.001, 
and 0.004. In Table~\ref{tab:core-collapse} we show the results for the fits using progenitor metallicities 
ranging from 0 to 0.02 (solar). The Z=0.001, Z=0.004, and Z=0.02 models do show a relatively small variation 
in $\chi^2$. The data are still compatible with a wide range of metallicities (0.001-0.02).

In Fig.~\ref{fig:bad_metal} we show the fit result for the Z=0.001 model. The main difference between this $Z$=0.001 model
and the $Z$=0.02 model is the amount of oxygen and neon produced. The neon peak is clearly less pronounced compared to the model
in Fig.~\ref{fig:badenes2}, while the plateau from silicon to nickel in the core-collapse contribution is nearly unaffected. 
That also confirms that the influence of metallicity differences in SNcc models on the Ar/Ca ratio is limited.

\subsubsection{Effect of IMF on core-collapse yields}

\begin{figure}[t]
\includegraphics[width=\columnwidth]{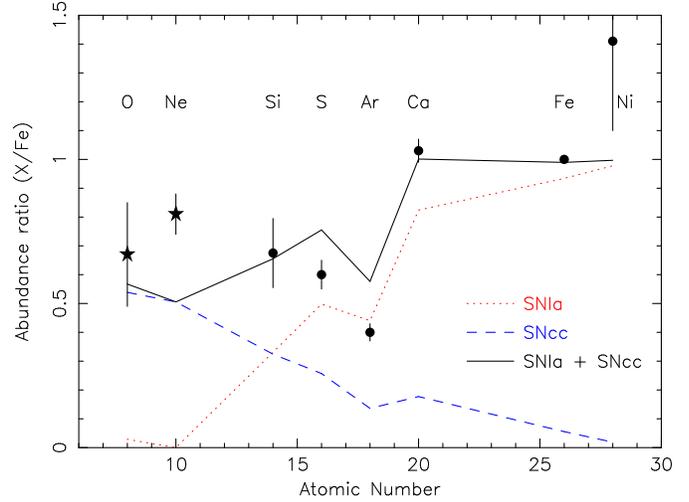}
\caption{Fit using the SNIa yields by \citet{badenes2006} and SNcc yields for Z=0.001 
metallicity progenitors \citep{nomoto2006} that are integrated over a ``top-heavy'' IMF.}
\label{fig:bad_top}
\end{figure}

We also fit the data for core-collapse models integrated over a ``top-heavy'' IMF ($x$=0), 
because presumably more high-mass stars form in low-metallicity environments.    
Table~\ref{tab:core-collapse} shows that the fits are similar to the fits using Salpeter IMF. 
In Fig.~\ref{fig:bad_top}, we show the fit 
for this metallicity. The main differences with Salpeter IMF models are in the O/Ne ratio.  
The abundance peak in the model is shifted from neon to oxygen with respect to the Salpeter 
models, which is less consistent with the neon abundance from RGS. Again, the    
plateau from silicon to nickel in the core-collapse model is barely affected. 

\subsection{The fraction of low mass stars that become Type Ia SNe}

This study indicates a much higher lifetime averaged ratio of type Ia to 
core-collapse supernovae than in the Galaxy.
The reason for this large contribution of SNIa is likely that
in late type galaxies, like our own galaxy, ongoing star formation ensures
an ongoing core collapse contribution. For clusters of galaxies, the 
star formation continued at a very reduced
level shortly after the formation of the cluster. This has some interesting
consequences. SNIa are likely the result of 
a thermonuclear runaway explosion of a C-O white dwarf in binary systems, 
caused by accretion from the secondary star. Since this involves
the formation of a white dwarf from a star with mass 
$M \lesssim 10~\mathrm{M}_{\sun}$,
there is a considerable delay between the period of star formation 
and the resulting SNIa explosion. In late type galaxies the subsequent
waves of star formation make it difficult to disentangle the SNIa
contribution from recent and old star formation periods. However,
cluster of galaxies are an interesting laboratory to study the fraction of
all stars that will eventually become type Ia supernovae, since the
star formation has shut down, and for the last few Gyr since formation,
the buffer of potential type Ia progenitors has nearly emptied.

As a result, the fraction of SNIa in clusters must be a good
approximation to the fraction of low mass stars that can become type Ia 
explosions. For a power law initial mass function we can write
\citep{yoshii96}:
\begin{equation}
\frac{SN Ia}{SN Ia + SN_{cc}} =
\frac{f_{SN Ia}\int_{M_{low}}^{M_{cc}} m^{-(1+x)} dm}
{f_{SN Ia}\int_{M_{low}}^{M_{cc}} m^{-(1+x)} dm + 
\int_{M_{cc}}^{M_{up}} m^{-(1+x)} dm},
\end{equation}
with $M_{low}$\ the lower limit to the mass of stars 
that can have contributed
to the type Ia production, and $M_{up}$\ the mass of the most massive stars.
The parameter of interest here is $f_{SN Ia}$, which is defined as the fraction of stars
with $M_{low} < M < M_{cc}$\ that will eventually explode as SNIa.

\begin{table}[t]
\caption{ The fraction, $f_{SN Ia}$, of low mass stars in the range $M_{low}$ to
$M_{cc}$ that will eventually result in SNIa.
It is based on our observed SNIa fraction of $\frac{SNIa}{SNIa+SNcc} =0.44\pm0.10$.
\label{typeIafractions}
}
\begin{center}
\begin{tabular}{lcc}
\hline\hline 
\noalign{\smallskip}
                & $M_{low} = 0.9~\mathrm{M}_{\sun}$  & {\bf $M_{low} = 1.5~\mathrm{M}_{\sun}$} \\
\noalign{\smallskip}
\hline
\noalign{\smallskip}
\multicolumn{3}{c}{Salpeter IMF ($x=1.35$)}\\
$M_{cc}=8~\mathrm{M}_{\sun}$  &   $4.1\pm1.7$\%  & {\bf $9\pm4$\% } \\
$M_{cc}=10~\mathrm{M}_{\sun}$ &   $2.9\pm0.1$\%  & {\bf $6\pm3$\% } \\
\noalign{\smallskip}
\multicolumn{3}{c}{\citet{kroupa02}
}\\{\smallskip}
$M_{cc}=8~\mathrm{M}_{\sun}$  &  $2.0\pm 0.8$\%  & {\bf $4.0\pm1.7$\% } \\
$M_{cc}=10~\mathrm{M}_{\sun}$ &   $1.3\pm 0.6$\%  & {\bf $2.7\pm1.1$\% } \\
\hline
\end{tabular}
\end{center}
\end{table}

It is clear that the absolute lower limit to $M_{low}$\ is the mass
of stars that can evolve to a C-O white dwarf within a Hubble time,
about 0.9 $\mathrm{M}_{\sun}$. However, in order to have  
sufficient mass
available in the binary to push the white dwarf over the Chandrasekhar
limit, it is usually assumed that the total mass of a binary producing a
SNIa explosion should exceed $\sim$ 3 $\mathrm{M}_{\sun}$
\citep{matteucci01}. This implies that the initial mass of the primary star
$\gtrsim$ 1.5 $\mathrm{M}_{\sun}$.

The time scale for producing a Type Ia supernova is most likely determined
by the evolution of the secondary, i.e. roughly by the duration of the
main sequence. If the secondary is 0.9 $\mathrm{M}_{\sun}$, this 
corresponds roughly to a Hubble time, but
there is considerable evidence that the mean delay time between star
formation and type Ia explosion is shorter, namely of the order of 1~Gyr.
\citet{strolger04} find a value of 2-4 Gyr. However, based on a lack of
observed SNIa in clusters of galaxies, \citet{maoz2004} find a 2$\sigma$
upper-limit of about 3 Gyr for the delay time. Very recently
\citet{mannucci06} argued for two channels for SNIa explosions, one with
a very short delay time of $10^8$~yr, and the other with 2-4 Gyr.
In other words, the majority of the type Ia explosions must occur
in binaries where the secondary star has a lifetime shorter than 2-4
Gyr. The mass of the secondary star that corresponds to such lifetimes
is about 1.25--1.5 $\mathrm{M}_{\sun}$ \citep{greggio83}. In the extreme
case, where the delay time is $\lesssim$ 2 Gyr, $M_{low}$ would be about
1.5 $\mathrm{M}_{\sun}$.

Table~\ref{typeIafractions} shows the values for  $f_{SN Ia}$\ using
different assumptions for $M_{low}$, $M_{cc}$ and the initial mass function.
We use the best fit SNIa rate of $0.44\pm 0.10$ (Table~\ref{tab:core-collapse}), 
but with slightly larger errors in order to allow for systematic uncertainties.
The values that we find for $M_{low}=1.5 \mathrm{M}_{\sun}$ are high compared 
to previous results based on other observational data, which suggests $f_{SN Ia}=1-5$\%
\citep{greggio83,yoshii96,matteucci01}. Note that these authors sometimes use
slightly different definitions for $M_{low}$ and $M_{cc}$.
Our lowest value is 1.3\%, which assumes $M_{cc}=10~\mathrm{M}_{\sun}$  and 
$M_{low}=0.9~\mathrm{M}_{\sun}$ together with the broken power law IMF of \citet{kroupa02}.
Notice also that $f_{SN Ia}$ refers to all stars with masses between $M_{low}$ and $M_{cc}$. 
We can estimate the probability that a binary produces a type Ia supernova from the value of $f_{SN Ia}$. 
We assume that roughly 50\% of all stars are formed in binaries, which introduces a factor of two. 
In addition, we need another factor of two because we want to count binaries instead of stars.  
Together, the probability that a binary produces a SNIa supernova is thus about four times higher then 
$f_{SN Ia}$.

Our derived supernova ratios therefore suggest that binary
systems in the appropriate mass range are very efficient in eventually 
forming SNIa explosions ($\sim$ 5--16\%, depending on the assumptions for the IMF 
and $M_{cc}$). We are aware that we ignore several complications in this simple calculation, 
such as an increased binarity fraction for massive stars that flattens the IMF 
for binary stars and binary mass ratios that may peak near 1. Also 
the instanteneous recycling approximation may introduce 
an additional uncertainty. However, a detailed calculation is beyond the scope 
of this paper.

\section{Conclusions}

We measure the abundances for silicon, sulfur, argon, calcium, iron, and nickel in a 
sample of clusters with XMM-Newton (EPIC), and we add a high-resolution
oxygen and neon measurement from RGS \citep{deplaa2006,werner2006}. From these data we conclude that:
\begin{itemize}
\item The Ar/Ca ratio in clusters is a good touchstone for determining the quality of type Ia models. 
The core-collapse contribution, which is about 50\% and not strongly dependent on the IMF or progenitor 
metallicity, does not have a significant impact on the Ar/Ca ratio.  
\item Current supernova type Ia models \citep{iwamoto1999}  
do not agree with our data, because they fail to produce the Ar/Ca and Ca/Fe abundance ratios. 
\item A major improvement of the supernova fits is obtained, when we use an empirically-modified 
supernova type Ia model, which is calibrated on the Tycho supernova remnant \citep{badenes2006}. This 
model largely solves the problems with the Ar/Ca and Ca/Fe abundance ratios by varying the density 
where the sound wave in the supernova turns into a shock and varying the ratio of the specific internal 
energies of ions and electrons at the shock.
\item The number ratio between supernova type Ia and core-collapse supernovae suggests that 
binary systems in the appropriate mass range are very efficient ($\sim$ 5--16\%) in eventually 
forming supernova type Ia explosions.
\item We find that the progenitors of the core-collapse supernovae which contributed to the ICM
abundances have probably been enriched. Progenitor abundances range from $Z\sim$0.001 to $Z\sim$0.02.
\item The intrinsic spread in abundance ratios between clusters is smaller than 30\%.
That means that the chemical histories of the clusters do not depend a lot on cluster 
temperature, temperature structure or merging activity.

\end{itemize}

\begin{acknowledgements}
We would like to thank Norbert Langer, Ton Raassen, Rob Izzard, and the anonymous referee for useful 
discussions. We are also grateful to Carles Badenes who kindly provided details about his work on the 
Tycho supernova remnant and to Steve Sembay who provided information about the current calibration 
status of the EPIC instruments. The work is based on observations obtained with XMM-Newton, an ESA 
science mission with instruments and contributions directly funded by ESA member states and the USA 
(NASA). The Netherlands Institute for Space Research (SRON) is supported financially by NWO, the 
Netherlands Organisation for Scientific Research.
\end{acknowledgements}

\bibliographystyle{aa}
\bibliography{clusters}

\appendix

\section{Abundance data}

In Table~\ref{tab:abundance} we list the abundances obtained from fits to the EPIC data. The MOS
and pn spectra are fitted simultaneously. We included a systematic error in the uncertainties on 
the Si/Fe and Ni/Fe abundance ratios (see Sect.~\ref{sec:cal} for
a discussion about systematic errors).    

\begin{table*}[t]
\caption{ Abundance ratios in the sample of clusters with respect to the solar abundances determined by 
\citet{lodders2003}. The listed ratio is calculated using $\frac{X/X_{\sun}}{Fe/Fe_{\sun}}$. 
We included a systematic error in the data points of Si/Fe and Ni/Fe. See Sect.~\ref{sec:cal} for
a discussion about systematic errors.}
\begin{center}
\begin{tabular}{l|ccccc|c}
\hline
Cluster &  Si/Fe	       & S/Fe	       & Ar/Fe         & Ca/Fe         & Ni/Fe         & Fe  \\  
\hline\hline
2A 0335 &  0.78  $\pm$ 0.09  &  0.636 $\pm$ 0.019  &  0.43 $\pm$ 0.04  &  0.95 $\pm$ 0.06   & 1.4 $\pm$ 0.4 &  0.741 $\pm$ 0.006 \\
A 85	&  0.72  $\pm$ 0.18   &  0.61  $\pm$ 0.15   &  0.4  $\pm$ 0.4  &  1.4  $\pm$ 0.4    & 1.0  $\pm$ 0.7  &  0.574 $\pm$ 0.018\\
A 133	&  0.64  $\pm$ 0.14   &  0.40  $\pm$ 0.09   &  0.6  $\pm$ 0.2  &  1.3  $\pm$ 0.3    & 1.6  $\pm$ 0.6  &  0.81  $\pm$ 0.02 \\
A 1651  &  0.0   $\pm$ 0.4    &  0.3   $\pm$ 0.4    &  0.0  $\pm$ 0.4  &  0.0  $\pm$ 0.6    & 1.4  $\pm$ 1.3  &  0.45  $\pm$ 0.03  \\
A 1689  &  0.3   $\pm$ 0.4    &  1.1   $\pm$ 0.4    &  0.8  $\pm$ 1.0  &  0.6  $\pm$ 0.9    & 0.6  $\pm$ 1.0  &  0.40  $\pm$ 0.02  \\
A 1775  &  0.57  $\pm$ 0.18   &  0.77  $\pm$ 0.14   &  0.5  $\pm$ 0.3  &  1.7  $\pm$ 0.4    & 1.6  $\pm$ 0.7  &  0.63  $\pm$ 0.02  \\
A 1795  &  0.75  $\pm$ 0.14   &  0.34  $\pm$ 0.08   &  0.2  $\pm$ 0.2  &  1.1  $\pm$ 0.3    & 1.3  $\pm$ 0.5  &  0.517 $\pm$ 0.009 \\
A 2029  &  0.4   $\pm$ 0.2    &  0.32  $\pm$ 0.17   &  0.00 $\pm$ 0.10  &  1.3  $\pm$ 0.5    & 1.6  $\pm$ 0.7  &  0.587 $\pm$ 0.017\\
A 2052  &  0.74  $\pm$ 0.12   &  0.76  $\pm$ 0.05   &  0.55 $\pm$ 0.11  &  1.35 $\pm$ 0.16   & 1.4 $\pm$ 0.4 &  0.682 $\pm$ 0.011 \\
A 2199  &  0.73  $\pm$ 0.13   &  0.60  $\pm$ 0.07   &  0.60 $\pm$ 0.16  &  1.04 $\pm$ 0.19   & 1.5  $\pm$ 0.5  &  0.532 $\pm$ 0.008\\
A 2204  &  0.75  $\pm$ 0.18   &  1.32  $\pm$ 0.19   &  0.0  $\pm$ 0.2  &  1.6  $\pm$ 0.6    & 1.6  $\pm$ 0.7  &  0.59  $\pm$ 0.02  \\
A 2589  &  0.55  $\pm$ 0.14   &  0.58  $\pm$ 0.10   &  0.4  $\pm$ 0.2  &  1.2  $\pm$ 0.3    & 1.8  $\pm$ 0.6  &  0.666 $\pm$ 0.019 \\
A 3112  &  0.70  $\pm$ 0.14   &  0.67  $\pm$ 0.09   &  0.5  $\pm$ 0.2  &  1.3  $\pm$ 0.3    & 1.9  $\pm$ 0.6  &  0.695 $\pm$ 0.016\\
A 3530  &  1.1   $\pm$ 0.6    &  1.1   $\pm$ 0.6    &  0.3  $\pm$ 0.9  &  1.2  $\pm$ 1.5    & 0.0  $\pm$ 2.0  &  0.28  $\pm$ 0.04 \\
A 3558  &  0.74  $\pm$ 0.16   &  0.41  $\pm$ 0.12   &  0.3  $\pm$ 0.3  &  1.6  $\pm$ 0.4    & 1.5  $\pm$ 0.6  &  0.478 $\pm$ 0.017 \\
A 3560  &  0.9   $\pm$ 0.3    &  0.4   $\pm$ 0.3    &  0.2  $\pm$ 0.5  &  1.0  $\pm$ 1.0    & 1.1  $\pm$ 1.3  &  0.39  $\pm$ 0.03  \\
A 3581  &  0.80  $\pm$ 0.10   &  0.75  $\pm$ 0.04   &  0.66 $\pm$ 0.09  &  1.32 $\pm$ 0.16   & 1.3 $\pm$ 0.4 &  0.654 $\pm$ 0.013 \\
A 3827  &  0.2   $\pm$ 0.3    &  0.0   $\pm$ 0.4    &  0.0  $\pm$ 0.7  &  2.9  $\pm$ 1.2    & 1.1  $\pm$ 1.2  &  0.38  $\pm$ 0.03 \\
A 3888  &  0.1   $\pm$ 0.5    &  1.8   $\pm$ 1.1    &  0.0  $\pm$ 1.3  &  1.0  $\pm$ 1.9    & 0.0  $\pm$ 1.8  &  0.30  $\pm$ 0.03 \\
A 4059  &  0.59  $\pm$ 0.13   &  0.45  $\pm$ 0.07   &  0.16 $\pm$ 0.17  &  0.7  $\pm$ 0.2    & 1.0  $\pm$ 0.5  &  0.687 $\pm$ 0.014 \\
MKW 3s  &  0.84  $\pm$ 0.13   &  0.63  $\pm$ 0.07   &  0.54 $\pm$ 0.18  &  1.2  $\pm$ 0.2    & 1.4  $\pm$ 0.5  &  0.551 $\pm$ 0.011 \\
S 159-03&  0.67  $\pm$ 0.10   &  0.55  $\pm$ 0.03   &  0.41 $\pm$ 0.07  &  0.87 $\pm$ 0.10   & 1.6 $\pm$ 0.4 &  0.533 $\pm$ 0.005 \\
\hline
\end{tabular}
\end{center}
\label{tab:abundance}
\end{table*}

\end{document}